\documentclass[a4paper,11pt]{article}
\usepackage{jheppub} 

\usepackage{graphicx}
\usepackage{subcaption}
\usepackage{caption}
\usepackage{enumitem}

\title{Higgs Production Classifier using Weak Supervision}

\author[a]{Kai-Feng Chen,}
\author[a]{Yi-An Chen,}
\author[a,b]{Cheng-Wei Chiang,}
\author[a]{and Feng-Yang Hsieh}

\affiliation[a]{Department of Physics, National Taiwan University, Taipei 10617, Taiwan}
\affiliation[b]{Physics Division, National Center for Theoretical Sciences, Taipei 10617, Taiwan}

\emailAdd{kfjack@phys.ntu.edu.tw}
\emailAdd{f08222011@ntu.edu.tw}
\emailAdd{chengwei@phys.ntu.edu.tw}
\emailAdd{f10222035@ntu.edu.tw}

\abstract{A reliable determination of the Higgs production mechanism in hadron collider experiments is essential in the program of the measurements of the Higgs couplings.  We employ weak supervision, CWoLa in particular, to train deep neural networks using real data of the diphoton events, in the hope of reducing biases resulting from Monte Carlo simulations. Models based on the convolutional neural network and the transformer are tested and compared. In particular, the classification performance gets slightly better when the photon information is removed from training on the low-luminosity region of $H \to \gamma \gamma$. We explicitly show that the performance can be improved when the training dataset is enlarged by data augmentation using physics-motivated methods.  We further demonstrate that the trained model can be successfully applied to the $H \to ZZ$ and $H \to Z\gamma$ events, showing that such classifiers are agnostic to Higgs decay modes provided they do not involve strong QCD corrections.}

\begin{document}

\maketitle
\flushbottom

\section{Introduction}
\label{sec:introduction}

    Since the discovery of the Higgs boson in 2012~\cite{ATLAS:2012yve,CMS:2012qbp}, much effort has been devoted to measuring its couplings and production mechanisms in detail~\cite{ATLAS:2022vkf,CMS:2022dwd}. Among all Standard Model (SM) Higgs production modes, gluon–gluon fusion (GGF) dominates the total cross-section~\cite{LHCHiggsCrossSectionWorkingGroup:2016ypw}. In contrast, the subdominant vector boson fusion (VBF) process provides a probe of the electroweak symmetry breaking and potential new physics effects~\cite{Greljo:2015sla}. Accurately separating VBF from the dominant GGF background is therefore essential for precision Higgs measurements and for enhancing the sensitivity of beyond-the-SM (BSM) searches.

    VBF–GGF discrimination typically relies on sequential kinematic cuts or multivariate classifiers such as boosted decision trees trained on high-level observables~\cite{ATLAS:2018jvf,CMS:2023tfj}. However, the increasing complexity of event topologies and the correlations among low-level features motivate the use of deep learning techniques~\cite{Feickert:2021ajf}. Deep neural networks (DNNs) have demonstrated superior performance in extracting global and local event features, such as convolutional neural networks (CNNs) that exploit spatial correlations in calorimeter or jet images~\cite{deOliveira:2015xxd,cnn2,Kasieczka:2017nvn,Macaluso:2018tck,cnn5,Li:2019ufu,Chung:2020ysf} and transformer-based models that capture long-range dependencies between particles or jets~\cite{Qu:2022mxj,Builtjes:2022usj,trans1,Spinner:2024hjm,Finke:2023veq,Li:2023xhj,Wang:2024rup}. Such approaches achieve significantly improved discriminative power compared to traditional cut-based or shallow learning methods~\cite{Chiang:2022lsn,Auricchio:2023syb}.

    Nevertheless, fully supervised learning strategies are not directly applicable in realistic collider analyses, as event-level truth labels (e.g., true production mode) are unavailable in experimental data. This limitation forces the use of simulated samples, which can introduce model dependencies and systematic uncertainties~\cite{DBLP:journals/corr/abs-1912-08001,Gavranovic:2023oam}. To overcome this challenge, we employ the \textit{Classification Without Labels} (CWoLa) framework~\cite{Metodiev:2017vrx}, which enables weakly supervised training using only mixed samples with different signal-to-background fractions. CWoLa has been successfully applied to jet tagging~\cite{Lee:2019ssx,Dolan:2023abg}, anomaly detection~\cite{Collins:2018epr,Collins:2019jip} and resonance searches~\cite{Beauchesne:2023vie}, but its application to Higgs production classification remains unexplored. Applying CWoLa to this task presents some challenges: the subtle kinematic differences between VBF and GGF further demand large training datasets for robust training, and limited statistics can degrade model performance in weakly supervised settings.

	To mitigate these limitations, we employ the data augmentation technique, which is effective in the CWoLa framework~\cite{Chen:2024nvc}. We introduce a physics-motivated data augmentation strategy based on the detector's azimuthal symmetry. By randomly rotating all event constituents along the azimuthal direction, we generate physically valid yet statistically independent variants of the same event. This augmentation effectively enhances the training statistics and reduces the fluctuation of weakly supervised learning, particularly in the low-luminosity region.

	Furthermore, we investigate the decay-mode agnosticism and transferability of CWoLa-trained classifiers. Since the Higgs boson is a color-singlet scalar, its decay products, particularly those carrying no color charge, are largely factorized from the VBF or GGF initial-state jets. It has been shown that in the supervised case, hadronic activity in an event can provide sufficient information to distinguish between VBF and GGF production~\cite{Chiang:2022lsn}. Building upon this, we test whether a CWoLa-trained model on $H \to \gamma\gamma$ events can be directly transferred to $H \to ZZ \to 4\ell$ and $H \to Z\gamma \to 2\ell\gamma$ events, combining weak supervision with transfer learning~\cite{Beauchesne:2023vie}. This approach further enhances data efficiency and tests the decay-channel independence of weakly supervised models.

	In summary, the goals of this work are threefold:
	\begin{enumerate}
        \item to evaluate the feasibility of applying the CWoLa framework to the classification of VBF and GGF Higgs production mechanisms using both CNN and transformer-based architectures;
        \item to quantify the improvement provided by azimuthal data augmentation in weakly supervised settings, particularly in the low-luminosity region;
        \item to assess the decay-mode transferability of CWoLa-trained classifiers across $H\to\gamma\gamma$, $H\to ZZ\to4\ell$, and $H \to Z\gamma \to 2\ell\gamma$ channels.
    \end{enumerate}
    These studies aim to establish a more data-efficient and decay-agnostic paradigm for weakly supervised Higgs production analyses at the CERN Large Hadron Collider (LHC). The analysis framework and all training scripts are publicly available to facilitate reproducibility and further studies\footnote{Codes are available at: \url{https://github.com/maplexgitx0302/NTUHEPML-CWoLa}. Datasets are available at Zenodo~\cite{hsieh_2025_17628988}}.

	The structure of this paper is organized as follows. Section~\ref{sec:methodology_weakly_supervised_learning_for_higgs_production} introduces the theoretical foundation of the CWoLa framework and outlines its implementation for distinguishing VBF and GGF Higgs production mechanisms, together with the neural network architectures and training strategies employed. Section~\ref{sec:experimental_setup_and_dataset_preparation} describes the Monte Carlo event generation, selection criteria, dataset construction, and preprocessing procedures, including the physics-inspired data augmentation. Section~\ref{sec:results_and_discussion} presents the classification results for the $H\to\gamma\gamma$ and $H\to ZZ\to4\ell$ channels, and investigates the transferability of models across decay modes, including the $H\to Z\gamma\to2\ell\gamma$ channel. Finally, section~\ref{sec:conclusions} summarizes the main findings and discusses the implications and prospects of applying weakly supervised learning to Higgs production studies.


\section{Methodology: weakly supervised learning for Higgs production}
\label{sec:methodology_weakly_supervised_learning_for_higgs_production}

    This section introduces the weakly supervised learning methodology adopted in this study, based on the CWoLa framework~\cite{Metodiev:2017vrx}. We first summarize the theoretical foundations of CWoLa and its relevance to collider data analysis. We then describe its implementation for distinguishing Higgs boson production mechanisms. Finally, we present the neural network architectures and training strategies used in this work.

    \subsection{Overview of CWoLa framework}
    \label{sub:overview_of_cwola_framework}

        In high-energy physics analyses, event-level truth labels that distinguish signal from background are typically inaccessible in experimental data. As a result, fully supervised learning techniques that rely on truth-level labels cannot be directly applied to real collider datasets. To address this limitation, we employ the CWoLa framework, a weakly supervised learning paradigm that enables the training of binary classifiers directly on partially labeled datasets.

        The CWoLa framework operates on two mixed datasets, each consisting of different and unknown proportions of signal and background events. A classifier is trained by full supervision to distinguish between these two mixed datasets by treating them as separate classes. At this stage, the classifier is only provided with the partial information of the dataset label on the events. Under the condition that the signal fractions of the two samples are not identical, the optimal CWoLa classifier is guaranteed to be a monotonic function of the truth likelihood ratio between the pure signal and background distributions. Consequently, in the asymptotic limit of infinite statistics and model capacity, the CWoLa-trained classifier is statistically equivalent to the fully supervised classifier.

        This property provides two key advantages: (1) it removes the need for event-level truth labels during training, and (2) it does not require prior knowledge of the signal-to-background ratios in the mixed samples, so long as they are different. Such features make CWoLa particularly well-suited for collider analyses, where signal or background regions can naturally provide samples with differing signal purities. In this study, we extend its use to the classification of Higgs boson production mechanisms, demonstrating that weak supervision can effectively separate VBF and GGF processes without relying on truth information.


    \subsection{Implementation for Higgs production classification}
    \label{sub:implementation_for_higgs_production_classification}

        The objective of this study is to construct a binary classifier distinguishing Higgs boson events produced via VBF and GGF processes. In this context, VBF events are treated as the \emph{signal} and GGF events as the \emph{background}. Because event-level truth labels are unavailable in experimental data, we adopt the CWoLa framework to train a classifier using only mixed samples. The central idea is that by learning to discriminate between events in different mixed samples, the model implicitly acquires the features that best separate pure VBF and GGF events.

        To emulate realistic experimental conditions, two mixed datasets are constructed from Monte Carlo simulations and denoted as the \emph{signal region} (SR) and the \emph{control region} (CR). The distinction is based on the parton flavor composition of the two leading reconstructed jets: events in which both jets originate from quarks are assigned to the SR, while events containing at least one gluon jet are assigned to the CR. The jet flavor information is obtained from the \verb|Delphes|~\cite{deFavereau:2013fsa} output, which preserves parton-level matching from event generation. Although such information is available only in simulation, in a realistic experimental setting, it could be inferred using auxiliary quark–gluon tagging algorithms~\cite{Gallicchio:2011xq}.

        The SR and CR datasets are divided into independent subsets for training and validation. The classifier is trained to distinguish SR from CR events using the CWoLa procedure, and the resulting model is evaluated on pure VBF and GGF samples to obtain an unbiased estimate of its discrimination performance. Further details on dataset composition, preprocessing, and data representation are provided in section~\ref{sec:experimental_setup_and_dataset_preparation}.

        Compared to traditional Higgs boson analyses based on the high-level observables such as invariant mass or angular separation~\cite{ATLAS:2020wny}, the CWoLa-based approach allows deep neural networks to exploit low-level information from particle-flow observables, learning correlations across full event topologies. This enables the discovery of subtle hadronic patterns that may be inaccessible to handcrafted selections. 


    \subsection{Neural network architectures and training strategy}
    \label{sub:neural_network_architectures_and_training_strategy}

        Two types of deep neural network architectures are employed under the CWoLa framework in our study: a CNN and a Particle Transformer (ParT). Both models are trained on identical datasets and under the same weak supervision conditions, differing only in their data representation and internal design.

        \paragraph{Convolutional Neural Network (CNN).}
        Each event is represented as an image defined on an $(\eta, \phi)$ grid, as detailed in section~\ref{sub:data_representation_for_neural_networks}. The CNN architecture follows the \textit{Event-CNN} design introduced in ref.~\cite{Chiang:2022lsn}, consisting of multiple convolutional layers with ReLU activations~\cite{DBLP:journals/corr/abs-1803-08375}, batch normalization~\cite{DBLP:journals/corr/IoffeS15}, and residual connections~\cite{DBLP:journals/corr/HeZRS15}, followed by fully connected layers and a sigmoid output node for binary classification. This architecture effectively captures spatial correlations in hadronic activity and has been optimized for event-level discrimination tasks. The total number of trainable parameters is approximately $2.7\times 10^5$. 

		\paragraph{Particle Transformer (ParT).}
		We employ the \textit{Particle Transformer} (ParT) architecture proposed in ref.~\cite{Qu:2022mxj}, designed to process unordered sets of event constituents. Unlike the original ParT configuration, our implementation omits the pairwise interaction matrix for two main reasons. First, the available inputs do not contain full four-momentum information for all objects, which limits the benefit of explicit interaction modeling. Second, our dataset consists of heterogeneous inputs—calorimeter towers, tracks, and decay products—rather than fully reconstructed particles, making it difficult to define consistent physical pairwise relations. Our implementation includes one particle attention block and one class attention block, providing a balance between expressive power and model stability. We found that the original ParT configuration tends to overfit with our datasets, while the simplified version yields more stable and reproducible results across repeated training trials. The model contains approximately $9.5\times10^3$ trainable parameters. A full list of hyperparameters is provided in Appendix~\ref{appendix:A}.

        \paragraph{Training setup.}
        All models are implemented using the \texttt{PyTorch 2.6.0} framework~\cite{DBLP:journals/corr/abs-1912-01703}, and optimized with the Adam optimizer~\cite{DBLP:journals/corr/KingmaB14}. The learning rates are set to $10^{-4}$ for the CNN and $4\times10^{-4}$ for the ParT model. Binary cross-entropy is used as the loss function, and the batch size is fixed at 512. Model performance is evaluated using the area under the receiver operating characteristic curve (AUC). Early stopping is applied by monitoring the validation AUC. The training is terminated after no improvement is observed for 10 epochs to prevent overfitting. Each reported result corresponds to an average over ten independent training runs with different random seeds, providing an estimate of statistical fluctuations.



\section{Experimental setup and dataset preparation}
\label{sec:experimental_setup_and_dataset_preparation}

    This section describes the simulation chain, dataset construction, and data processing pipeline used for this study. We first outline the event generation and selection procedures used to produce Monte Carlo samples. We then explain how mixed datasets are constructed for weak supervision, alongside an auxiliary fully supervised dataset for reference. Subsequently, we detail the event preprocessing, normalization, and data representation schemes adopted for convolutional and transformer-based neural networks.  Finally, we present the physics-inspired data augmentation technique designed to enhance training statistics and model stability.

    \subsection{Event generation and selection}
    \label{sub:event_generation_and_selection}

        SM Higgs boson events produced via VBF and GGF processes are simulated for the LHC at a center-of-mass energy of $\sqrt{s} = 14$~TeV. Event generation is performed using \verb|MadGraph 3.3.1|~\cite{Alwall:2014hca} with the \verb|NNPDF23_nlo_as_0119| PDF set~\cite{Ball:2012cx} for both production modes, with Higgs decays into $H \to \gamma\gamma$, $H \to ZZ \to 4\ell$, and $H \to Z\gamma \to 2\ell\gamma$. The parton showering and hadronization are simulated by \verb|Pythia 8.306|~\cite{Sjostrand:2014zea} with \verb|NNPDF2.3 LO| PDF set. The detector simulation is conducted by \verb|Delphes 3.4.2|~\cite{deFavereau:2013fsa} with the default CMS detector card. Jet reconstruction is carried out using \verb|FastJet 3.3.2|~\cite{Cacciari:2011ma} with the anti-$k_t$ algorithm~\cite{Cacciari:2008gp} and a jet radius parameter of $R = 0.4$. Jets are required to have transverse momentum $p_{\mathrm{T}} > 25$~GeV.

		After detector simulation, only events containing at least two reconstructed jets and the corresponding Higgs decay products are retained. For the $H \to \gamma\gamma$ channel, two photons are required, and the invariant mass of the two leading photons must satisfy $120~\mathrm{GeV} \le m_{\gamma\gamma} \le 130~\mathrm{GeV}$. For the $H \to ZZ \to 4\ell$ channel, events are required to contain at least four reconstructed leptons. For the $H \to Z\gamma \to 2\ell\gamma$ channel, events are required to contain at least one photon and two reconstructed leptons. The selection efficiencies and cumulative passing rates for each cut are summarized in table~\ref{tab:cutflow}.

		\begin{table}[htpb]
			\centering
			\caption{Cutflow table for the VBF and GGF production in the $H \to \gamma\gamma$, $H \to ZZ \to 4\ell$, and $H \to Z\gamma \to 2\ell\gamma$ channels. The ``Efficiency'' denotes the fraction of events passing a given cut relative to the previous step, while the ``Passing rate'' represents the cumulative fraction of events surviving all cuts up to that stage, relative to the initial number of events. We only impose the Higgs decay-product invariant mass cut in the $H \to \gamma\gamma$ case; otherwise, the number of passing events will be even more limited in the cases of the $ZZ$ and $Z\gamma$ modes.}
			\label{tab:cutflow}
			\subfloat[$H \to \gamma\gamma$]{
				\begin{tabular}{c|cc|cc}
																   & \multicolumn{2}{c|}{VBF}   & \multicolumn{2}{c}{GGF}  \\
					Cut                                            & Efficiency & Passing rate & Efficiency & Passing rate \\ \hline
					$n_{\gamma} \ge 2$                             & 0.531      & 0.531        & 0.483      & 0.483        \\
					$n_j \ge 2$                                    & 0.807      & 0.429        & 0.193      & 0.093        \\
					$m_{\gamma\gamma} \in [120, 130]~\mathrm{GeV}$ & 0.949      & 0.407        & 0.953      & 0.088        
				\end{tabular}
			} \\
			\bigskip
            \subfloat[$H \to ZZ \to 4\ell$]{
				\begin{tabular}{c|cc|cc}
									 & \multicolumn{2}{c|}{VBF}  & \multicolumn{2}{c}{GGF}  \\
					Cut              & Efficiency & Passing rate & Efficiency & Passing rate \\ \hline
					$n_{\ell} \ge 4$ & 0.190      & 0.190        & 0.273      & 0.273        \\
					$n_j \ge 2$      & 0.868      & 0.165        & 0.252      & 0.069        
				\end{tabular}
			} \\
            \bigskip
			\subfloat[$H \to Z\gamma \to 2\ell\gamma$]{
				\begin{tabular}{c|cc|cc}
									   & \multicolumn{2}{c|}{VBF}  & \multicolumn{2}{c}{GGF}   \\
					Cut                & Efficiency & Passing rate & Efficiency & Passing rate \\ \hline
					$n_{\gamma} \ge 1$ & 0.635      & 0.635        & 0.582      & 0.582        \\
					$n_{\ell} \ge 2$   & 0.477      & 0.303        & 0.498      & 0.290        \\
					$n_j \ge 2$        & 0.792      & 0.240        & 0.172      & 0.050       
				\end{tabular}
			}
		\end{table}

		For each simulated event, particle–flow observables (transverse momentum $p_{\mathrm{T}}$, pseudorapidity $\eta$, and azimuthal angle $\phi$) are recorded for calorimeter towers and reconstructed tracks. Additionally, the four-momenta of Higgs decay products are retained for subsequent analysis.


    \subsection{Construction of mixed and labeled datasets}
    \label{sub:construction_of_mixed_and_labeled_datasets}

       \subsubsection*{Mixed datasets for CWoLa}

        To realize the CWoLa framework, SR and CR datasets are constructed from Monte Carlo simulations and used as pseudo-experimental data. The separation between SR and CR follows the definition based on the parton flavor composition of the two leading jets, as described in section~\ref{sub:implementation_for_higgs_production_classification}. 

\begin{figure}[htbp]
            \centering
        
            \begin{subfigure}[b]{0.32\textwidth}
                \centering
                \includegraphics[width=\textwidth]{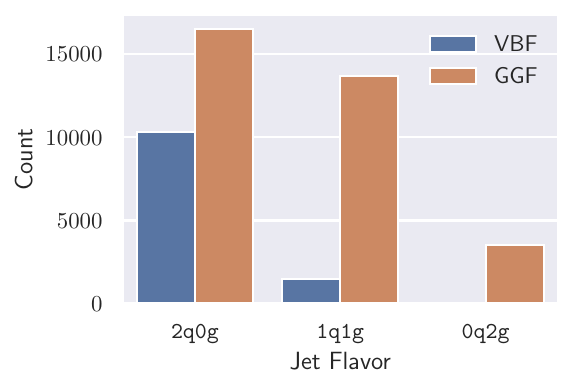}
                \caption{$H\to\gamma\gamma$}
                \label{fig:jet-flavor_diphoton}
            \end{subfigure}
            \hfill
            \begin{subfigure}[b]{0.32\textwidth}
                \centering
                \includegraphics[width=\textwidth]{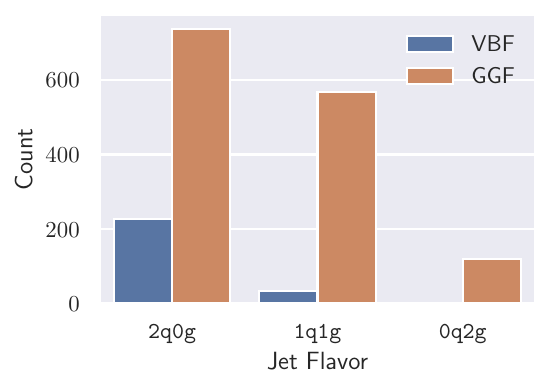}
                \caption{$H\to ZZ\to 4\ell$}
                \label{fig:jet-flavor_zz4l}
            \end{subfigure}
            \hfill
            \begin{subfigure}[b]{0.32\textwidth}
                \centering
                \includegraphics[width=\textwidth]{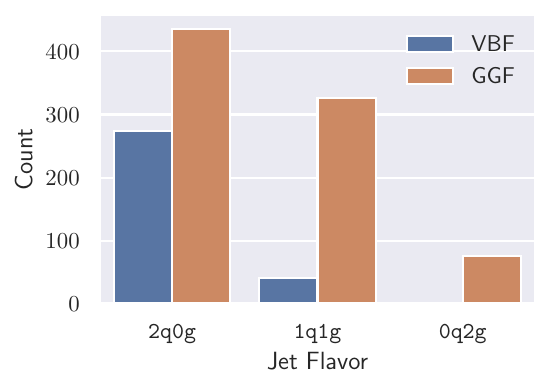}
                \caption{$H\to Z\gamma\to 2\ell\gamma$}
                \label{fig:jet-flavor_za2l}
            \end{subfigure}
        
            \caption{Distribution of jet flavor compositions at $\mathcal{L} = 3000~\mathrm{fb}^{-1}$ for $H\to\gamma\gamma$ and $H\to ZZ\to 4\ell$. The categories \texttt{2q0g}, \texttt{1q1g}, and \texttt{0q2g} correspond to events with two quark-initiated jets, one quark and one gluon jet, and two gluon-initiated jets, respectively. The SR is defined by the \texttt{2q0g} category, while the CR includes both \texttt{1q1g} and \texttt{0q2g} events.}
            \label{fig:jet-flavor}
        \end{figure}

		The corresponding SM branching ratios are $\mathrm{BR}(H \to \gamma\gamma) = 2.27 \times 10^{-3}$, $\mathrm{BR}(H \to 4\ell, \ell = e, \mu) = 1.24 \times 10^{-4}$, and $\mathrm{BR}(H \to Z\gamma \to 2\ell\gamma, \ell = e, \mu) = 1.03 \times 10^{-4}$; the Higgs production cross sections at $\sqrt{s} = 14~\mathrm{TeV}$ are $\sigma_{\mathrm{VBF}} = 4.278~\mathrm{pb}$ and $\sigma_{\mathrm{GGF}} = 54.67~\mathrm{pb}$~\cite{LHCHiggsCrossSectionWorkingGroup:2016ypw}. Using these values and the selection efficiencies summarized in table~\ref{tab:cutflow}, the expected numbers of events in the SR and CR can be estimated for a given integrated luminosity. Table~\ref{tab:event_numbers} shows the event yields at $\mathcal{L} = 3000~\mathrm{fb}^{-1}$, and the respective distributions of each jet flavor count are shown in figure~\ref{fig:jet-flavor}.

		\begin{table}[htpb]
			\centering
			\caption{Expected numbers of events in each category at $\mathcal{L} = 3000~\mathrm{fb}^{-1}$.}
			\label{tab:event_numbers}
			\subfloat[$H \to \gamma\gamma$]{
				\begin{tabular}{c|cc}
					& VBF & GGF \\ \hline
					SR & 10229 & 16828 \\
					CR & 1596  & 16865 \\
				\end{tabular}
			} 
			\quad
			\subfloat[$H \to ZZ \to 4\ell$]{
				\begin{tabular}{c|cc}
					& VBF & GGF \\ \hline
					SR & 228 & 722 \\
					CR & 34  & 704 \\
				\end{tabular}
			}
            \quad
			\subfloat[$H \to Z\gamma \to 2\ell\gamma$]{
				\begin{tabular}{c|cc}
					& VBF & GGF \\ \hline
					SR & 275 & 402 \\
					CR & 42  & 437 \\
				\end{tabular}
			}
		\end{table}

		Among these samples, 80\% of the events are used for training and 20\% for validation. An additional 10k VBF and 10k GGF events are reserved for testing. To evaluate the robustness of the neural network training, larger datasets are prepared, from which distinct mixed samples are resampled for each independent training instance. This approach ensures statistical independence across trainings and provides a reliable estimate of performance fluctuations in the datasets. 

        \subsubsection*{Datasets for fully supervised learning}
    
        To estimate the upper bound of classification performance achievable under ideal labeling, we construct fully supervised datasets where each event carries its truth label (VBF or GGF). For each production mode, 100k, 25k, and 25k events are used for training, validation, and testing, respectively. These datasets serve as a benchmark for assessing the relative performance of weakly supervised learning.


	\subsection{Data preprocessing}
	\label{sub:data_preprocessing}

		Before constructing the image and sequence representations for training, each event undergoes the following preprocessing steps:
		\begin{enumerate}[label=\Roman*.]
			\item The variance of $\phi$ of all constituents in an event is first computed. If this variance exceeds 0.5, all $\phi$ values are shifted by $\pi$ to center the distribution. This procedure prevents a significant fraction of event constituents from crossing the $\pm\pi$ boundary.
			
			\item The $\phi$ coordinates of all event constituents are then centered with respect to the $p_{\mathrm{T}}$-weighted mean:
				\[
					\phi \to \phi - \frac{1}{p_{\mathrm{T}}^{\mathrm{sum}}} \sum_i p_{\mathrm{T}, i}\phi_i, \quad p_{\mathrm{T}}^{\mathrm{sum}} = \sum_i p_{\mathrm{T}, i},
				\]
				where $p_{\mathrm{T}, i}$ and $\phi_i$ denote the transverse momentum and azimuthal angle of the $i$-th constituent, respectively. This step ensures that the event is rotationally aligned around its $p_{\mathrm{T}}$-weighted centroid. 
			
			\item Events are divided into four $(\eta,\phi)$ quadrants. The quadrant with the highest total transverse momentum is identified and reflected into the $(\phi > 0, \eta > 0)$ region by mirroring along the $\phi = 0$ and $\eta = 0$ axes.
		\end{enumerate}

    
    \subsection{Data representation for neural networks}
    \label{sub:data_representation_for_neural_networks}

        The representation of event data depends on the neural network architecture employed, as alluded to in section~\ref{sub:neural_network_architectures_and_training_strategy}. Two complementary formats are adopted: an image-based representation for convolutional networks and a set-based representation for transformer architectures.

        \subsubsection*{Image-based representation (for CNNs)}

		Each event is converted into a three-channel image defined on a $40\times40$ grid covering $\phi \in [-\pi, \pi]$ and $\eta \in [-5, 5]$. The three channels correspond to calorimeter towers, tracks, and Higgs decay products, respectively. It should be noted that the tower and track channels may also contain constituents from the Higgs decay products.

\begin{figure}[htbp]
			\centering
			\includegraphics[width=\textwidth]{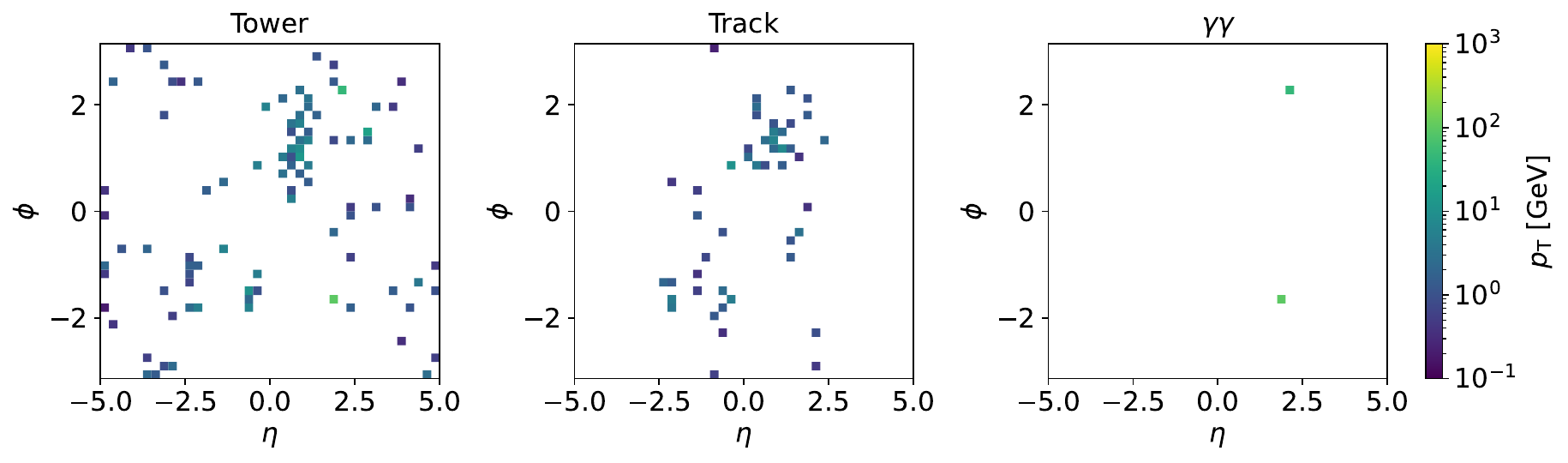}
			\caption{The GGF processes with $H\to\gamma\gamma$ in the image-based representation. Each event image consists of three channels corresponding to tower, track, and decay-product information. Note that tower and track channels may also contain constituents from the Higgs decay products.}
			\label{fig:pt_comparison}
		\end{figure}

		When excluding decay-product information, two operations are performed: (1) the decay-product channel is removed, and (2) the pixel values corresponding to decay-product positions in the remaining channels are set to zero. While this procedure may also remove contributions from other particles that fall into the same pixel, it affects only a very small fraction of pixels and does not alter the overall event structure. Compared with simply discarding the decay-product channel, this refined removal procedure has a negligible effect on the $H \to \gamma\gamma$ dataset but significantly impacts the model's generalization when applied to other channels. Figure~\ref{fig:pt_comparison} illustrates the $p_{\mathrm{T}}$ distributions in this image-based representation of a GGF event, with the Higgs boson decaying into diphotons.

        \subsubsection*{Set-based representation (for ParTs)}

		For set-based architectures such as transformers, each event is represented as a collection of reconstructed objects or particles. Each object is described by a six-dimensional feature vector. The first three components correspond to the kinematic variables $(p_{\mathrm{T}}, \eta, \phi)$, while the remaining components are the one-hot encoded type identifiers (tower, track, or decay product). This representation allows the model to process heterogeneous object types within a unified feature space.

		To remove decay-product information, all particles $i$ are excluded if their angular separation from a decay product satisfies $|\Delta\phi| < \pi/40$ and $|\Delta\eta| < 5/40$, where $\Delta\phi\equiv \phi_i-\phi_{\text{decay}}$ and $\Delta\eta\equiv\eta_i-\eta_{\text{decay}}$. These thresholds are chosen to match the corresponding $\phi$ and $\eta$ grid divisions of the image representation, ensuring consistency between the image-based and set-based removal procedures.

        \subsubsection*{$p_\mathrm{T}$ normalization}
        
        To mitigate the potential sculpting effect in CWoLa training, $p_{\mathrm{T}}$ normalization is applied to both image and sequence data. Following the approach of~\cite{Chen:2024nvc}, each event image (or set) is standardized by subtracting its mean pixel (constituent) $p_{\mathrm{T}}$ and dividing by the corresponding standard deviation. This normalization ensures that the learning process focuses on event topology rather than absolute energy scales.

	
	\subsection{Physics-inspired data augmentation}
	\label{sub:physics_inspired_data_augmentation}

		To address the data scarcity issue in training sets, we employ a physics-inspired data augmentation technique based on azimuthal symmetry, referred to as $\phi$-shifting. Due to the cylindrical symmetry of the LHC detectors, rotating all event constituents by a common azimuthal offset does not alter the underlying physics of the event. This property allows the generation of additional statistically independent samples without modifying the event kinematics or topology.

		After event preprocessing, a random rotation angle $\theta$ is chosen uniformly from the range $[-\pi, \pi]$. Each constituent in the event is then transformed according to
		\[
			\phi_i \rightarrow \phi_i + \theta,
		\]
		where $\phi_i$ denotes the azimuthal angle of the $i$-th object. To ensure the resulting values remain within the detector range $[-\pi, \pi]$, the shifted angles could be further moved by $2\pi$. Other kinematic quantities, such as transverse momentum $p_\text{T}$, pseudorapidity $\eta$, and the particle type, remain unchanged\footnote{We also studied the $\eta-\phi$ smearing and $p_{\text{T}}$ smearing methods, but we did not observe noticeable improvement.}.

		To enhance dataset diversity, we use a random, distinct shifting angle for each event and generate multiple augmented versions of each event by repeating this process with various random angles. In our experiments, we typically use fivefold (+5) and tenfold (+10) enlargements of the training sample. The augmentation is applied before data representation, ensuring consistency between image-based (CNN) and set-based (Transformer) representations.



\section{Results and discussions}
\label{sec:results_and_discussion}

    This section presents an evaluation of Higgs production classification within the CWoLa framework. We investigate three aspects: (1) the baseline discrimination performance using $H\to\gamma\gamma$ events, (2) the effect of azimuthal data augmentation on training stability and efficiency, and (3) the transferability of trained models across different Higgs decay channels. The training datasets correspond to the following integrated luminosities:
    \[
        \mathcal{L} \in \{100,\, 300,\, 900,\, 1800,\, 3000\}~\mathrm{fb^{-1}} ,
    \]
    unless otherwise specified. Furthermore, to mitigate the data scarcity, $\phi$-shifting augmentation is applied by shifting the $\phi$ coordinates of all constituents. Details of the data preparation, augmentation strategy, and experimental setup are already provided in section~\ref{sec:experimental_setup_and_dataset_preparation}.

    All performance metrics are reported in terms of the AUC, evaluated on pure VBF and GGF testing samples for an unbiased measure of discrimination. Each experiment is repeated over ten random seeds to estimate statistical fluctuations, and the mean values with one standard deviation are presented as shaded bands in the figures.

    \subsection{Baseline training on \texorpdfstring{$H \to \gamma\gamma$}{H to diphoton}}
    \label{sub:baseline_training_on_diphoton}

\begin{figure}[htbp]
            \centering
            \begin{subfigure}[b]{0.475\textwidth}
                \centering
                \includegraphics[width=\textwidth]{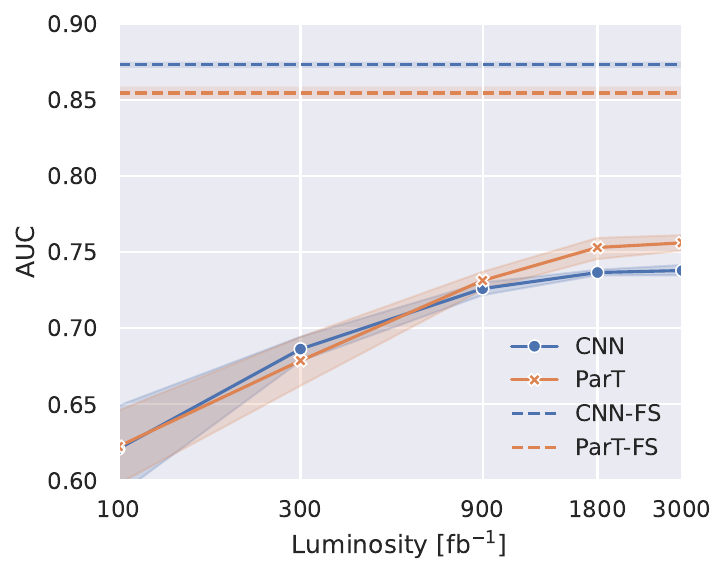}
                \caption{With photon information}
                \label{fig:CNN_ParT_w_photon}
            \end{subfigure}
            \hfill
            \begin{subfigure}[b]{0.475\textwidth}
                \centering
                \includegraphics[width=\textwidth]{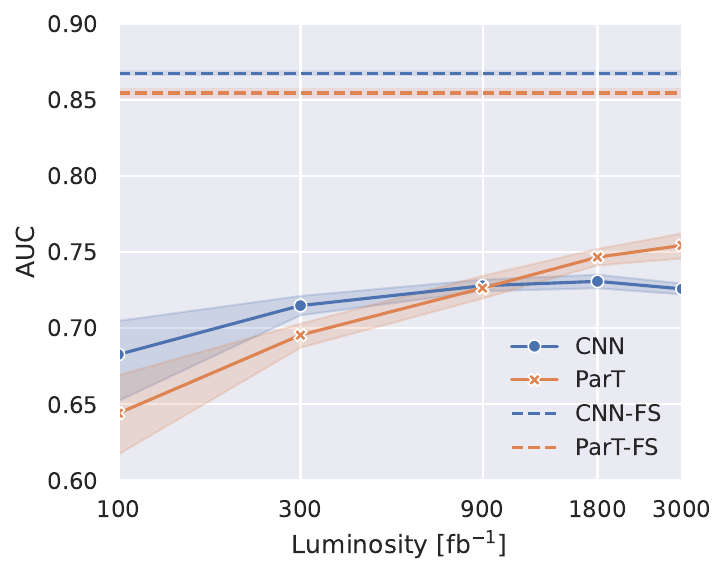}
                \caption{Without photon information}
                \label{fig:CNN_ParT_wo_photon}
            \end{subfigure}
        
            \caption{Test AUC of CNN and ParT models trained on $H \to \gamma\gamma$ events. Results are shown (a) with and (b) without decay-product information. Fully supervised (FS) baselines are included for reference. Each point represents the mean AUC over ten training seeds, and the shaded bands indicate one standard deviation.}
            \label{fig:diphoton_performance}
        \end{figure}

        The $H\to\gamma\gamma$ channel serves as the benchmark for evaluating the weakly supervised training. It offers a larger event yield and a simpler topology compared to the $H\to ZZ\to 4\ell$ and $H \to Z\gamma \to 2\ell\gamma$ channels, enabling robust statistical comparisons between architectures and data representations. Figure~\ref{fig:diphoton_performance} summarizes the AUC values obtained from CNN and ParT models trained on $H\to\gamma\gamma$ events, both with and without including the decay-product (photon) information.

		Figure~\ref{fig:CNN_ParT_w_photon} shows the comparison between CNN and ParT trained on the datasets with photon features. As expected, the neural network performance improves and the training fluctuations decrease as the size of the training dataset increases. When the integrated luminosity is below $900~\mathrm{fb}^{-1}$, both CNN and ParT models achieve comparable AUC values. At higher luminosities, ParT consistently outperforms CNN, demonstrating its stronger capacity to exploit complex correlations when a sufficient number of training events is available.

		Figure~\ref{fig:CNN_ParT_wo_photon} shows the performance of the models trained on the photon information-removed datasets. Remarkably, excluding the photon information does not degrade the network performance, suggesting that both architectures can effectively discriminate VBF and GGF processes based solely on hadronic activity. At high luminosities, the model performance with and without photon information becomes nearly identical, indicating that photon-related features contribute only marginally to the discrimination task. The hadronic activity alone provides sufficient information to distinguish between VBF and GGF events. 

        Interestingly, at lower luminosities, models trained without photon information achieve higher AUC values. This behavior suggests that, when the available training data are limited, the networks tend to overfit to the simpler photon features rather than learning the more complex hadronic patterns. When photon information is removed, the models are forced to focus on hadronic patterns, thereby achieving better generalization and higher performance in the low-statistics regime. These results demonstrate that the primary discriminative power arises from the hadronic activity in the events, even in the weakly supervised framework, and that explicit inclusion of photon information is not essential for achieving optimal classification performance\footnote{The same conclusion in the fully supervised learning scenario had already been noted in Ref.~\cite{Chiang:2022lsn}.}.

        For comparison, we also include the results from fully supervised (FS) training using the labeled datasets described in section~\ref{sub:construction_of_mixed_and_labeled_datasets}. Since the FS classifier represents the optimal performance achievable with truth labels, it is trained on a fixed number dataset rather than luminosity-dependent mixtures. As a result, its performance appears constant across the luminosity axis in our plots. The FS provides the upper bound for the weakly supervised CWoLa classifiers. The gap between the two indicates substantial room for improvement.


    \subsection{Impact of data augmentation}
    \label{sub:impact_of_data_augmentation}

\begin{figure}[htbp]
            \centering
        
            \begin{subfigure}[b]{0.475\textwidth}
                \centering
                \includegraphics[width=\textwidth]{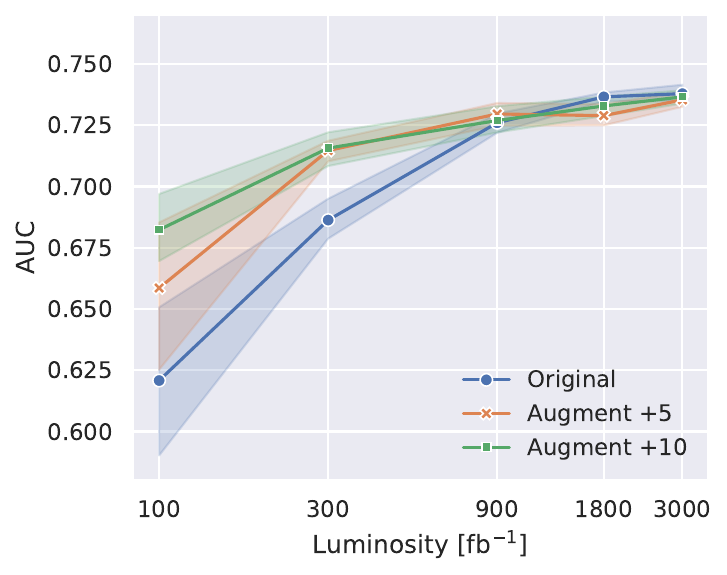}
                \caption{CNN with photon information}
            \end{subfigure}
            \hfill
            \begin{subfigure}[b]{0.475\textwidth}
                \centering
                \includegraphics[width=\textwidth]{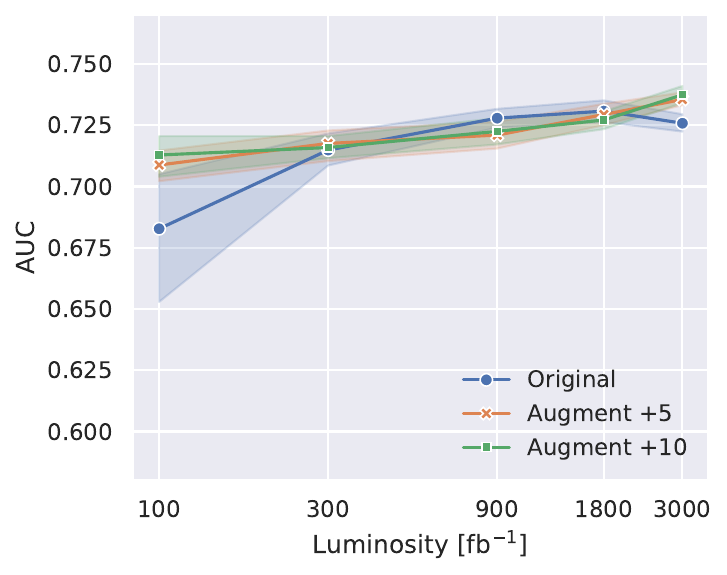}
                \caption{CNN without photon information}
                \label{fig:result_diphoton_aug-exCNN}
            \end{subfigure}
            \\[0.75cm]
            \begin{subfigure}[b]{0.475\textwidth}
                \centering
                \includegraphics[width=\textwidth]{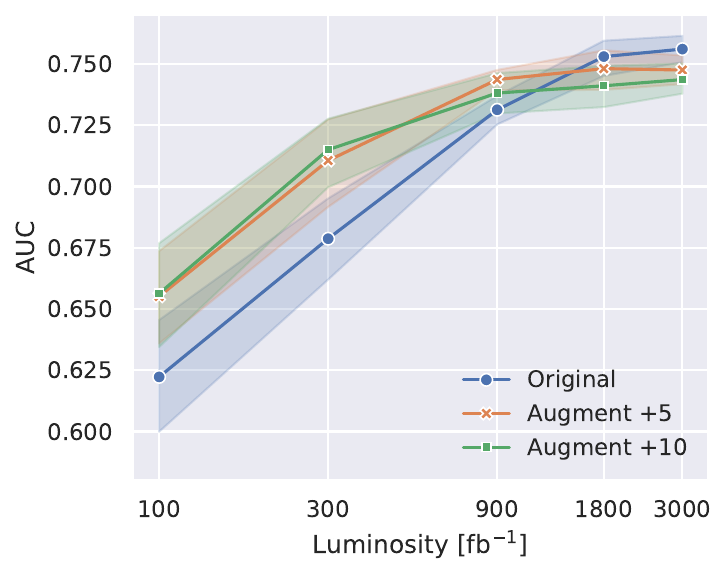}
                \caption{ParT with photon information}
            \end{subfigure}
            \hfill
            \begin{subfigure}[b]{0.475\textwidth}
                \centering
                \includegraphics[width=\textwidth]{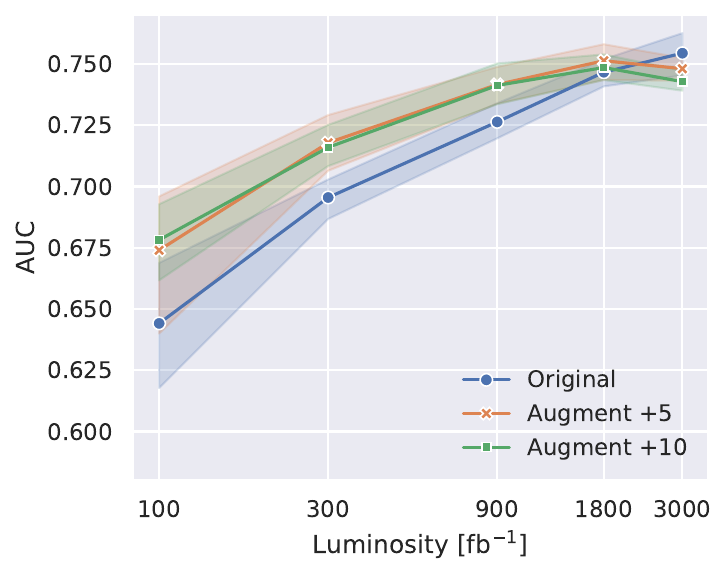}
                \caption{ParT without photon information}
                \label{fig:result_diphoton_aug-exParT}
            \end{subfigure}
        
            \caption{Effect of $\phi$-shifting augmentation on $H \to \gamma\gamma$ classification performance. Each point represents the mean AUC over ten training seeds, with shaded bands indicating one standard deviation.}
			\label{fig:result_diphoton_aug}
        \end{figure}

        As shown in figure~\ref{fig:diphoton_performance}, the neural network performance at low luminosities is unsatisfactory, primarily due to the limited number of training events. To mitigate this data scarcity, we employ data augmentation techniques, which have been shown to enhance training stability and performance in the CWoLa framework~\cite{Chen:2024nvc}. In particular, we apply the $\phi$-shifting augmentation to increase the effective size and diversity of the training samples. Figure~\ref{fig:result_diphoton_aug} presents the AUC values for the CNN and ParT models trained on $H \to \gamma\gamma$ datasets with fivefold (+5) and tenfold (+10) augmentation levels.

		For CNNs, the $\phi$-shifting augmentation substantially improves both the mean AUC and training stability in the low-luminosity regime ($\mathcal{L} \lesssim 300~\mathrm{fb}^{-1}$). Beyond $\mathcal{L} = 900~\mathrm{fb}^{-1}$, the improvement saturates, and the augmented and baseline performances converge. The benefit is more pronounced when photon information is included, suggesting that augmentation helps mitigate overfitting to the simpler photon features in low-statistics conditions and encourages the model to learn more robust hadronic correlations.

        For ParT models, similar trends are observed: augmentation enhances performance up to $\mathcal{L} \approx 900~\mathrm{fb}^{-1}$, with fivefold replication already sufficient. Increasing to tenfold offers no further gain and can even lead to mild over-regularization at the highest luminosities. Overall, the $\phi$-shifting augmentation provides a statistically efficient and physically consistent means of improving weakly supervised training, particularly when the available data are limited.


    \subsection{Performance on \texorpdfstring{$H \to ZZ \to 4\ell$}{H -> ZZ -> 4l}}
	\label{sub:performance_on_h_to_zz_to_4l}
		
		To investigate whether the models are agnostic to the Higgs decay mode and to establish a baseline for transferability studies, we perform CWoLa training on the $H \to ZZ \to 4\ell$ decay channel. Due to the small branching ratio $\mathrm{BR}(H \to 4\ell) = 1.24 \times 10^{-4}$ and the resulting limited event yield even at $\mathcal{L} = 3000~\mathrm{fb}^{-1}$ (see table~\ref{tab:event_numbers}), we include larger luminosities for the CWoLa training:
		\[
			\mathcal{L} \in \{9000,\, 18000,\, 30000\}~\mathrm{fb}^{-1}.
		\]

\begin{figure}[htbp]
            \centering
        
            \begin{subfigure}[b]{0.475\textwidth}
                \centering
                \includegraphics[width=\textwidth]{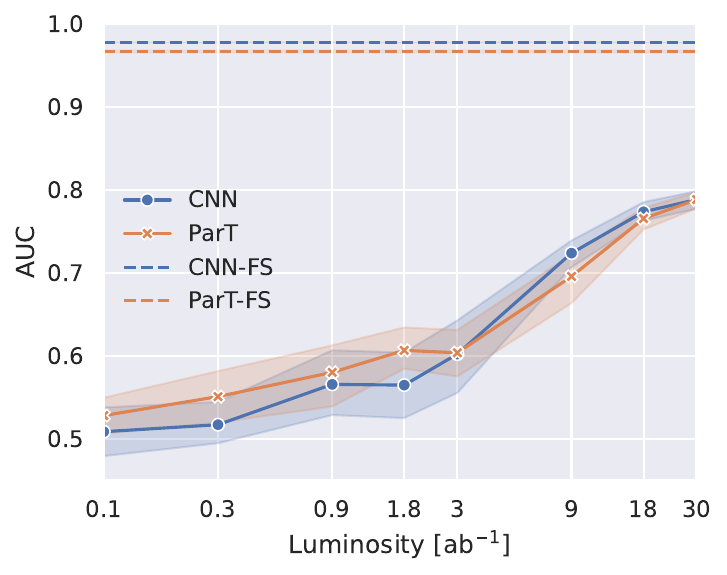}
                \caption{With lepton information}
                \label{fig:CNN_ParT_w_lepton}
            \end{subfigure}
            \hfill
            \begin{subfigure}[b]{0.475\textwidth}
                \centering
                \includegraphics[width=\textwidth]{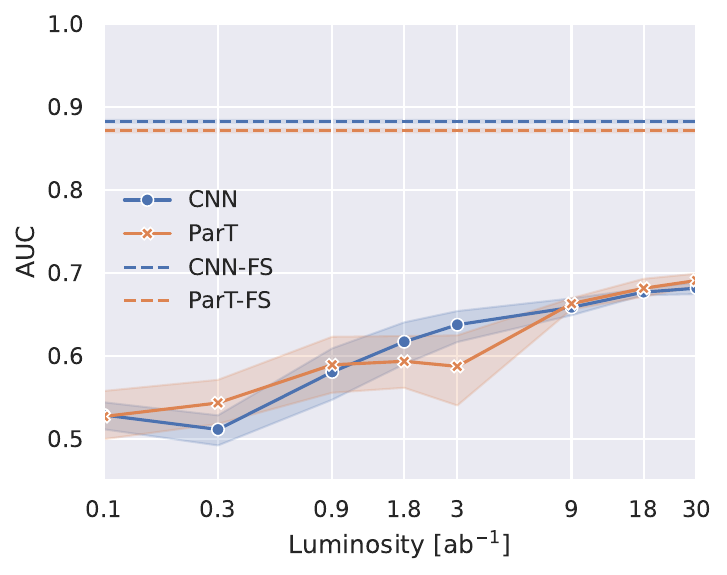}
                \caption{Without lepton information}
                \label{fig:CNN_ParT_wo_lepton}
            \end{subfigure}
        
            \caption{Test AUC of CNN and ParT models trained on $H \to ZZ \to 4\ell$ events, evaluated (a) with and (b) without decay-product information. The ``FS'' stands for fully supervised learning. Each point represents the mean AUC over ten training seeds, and the shaded bands indicate one standard deviation.}
    		\label{fig:result_zz4l}
        \end{figure}

		Figure~\ref{fig:result_zz4l} shows the AUC performance of CNN and ParT models trained on $H \to ZZ \to 4\ell$ samples, with and without the inclusion of decay-product (lepton) information. As expected, the model performance improves gradually with increasing training luminosity. However, even at $\mathcal{L} = 3000~\mathrm{fb}^{-1}$, the achieved AUC values remain modest and the training fluctuations are substantial, reflecting the severe data scarcity in this decay channel. The luminosity levels required to reach stable training are well beyond those achievable in realistic experimental conditions.

\begin{figure}[htbp]
            \centering
        
            \begin{subfigure}[b]{0.475\textwidth}
                \centering
                \includegraphics[width=\textwidth]{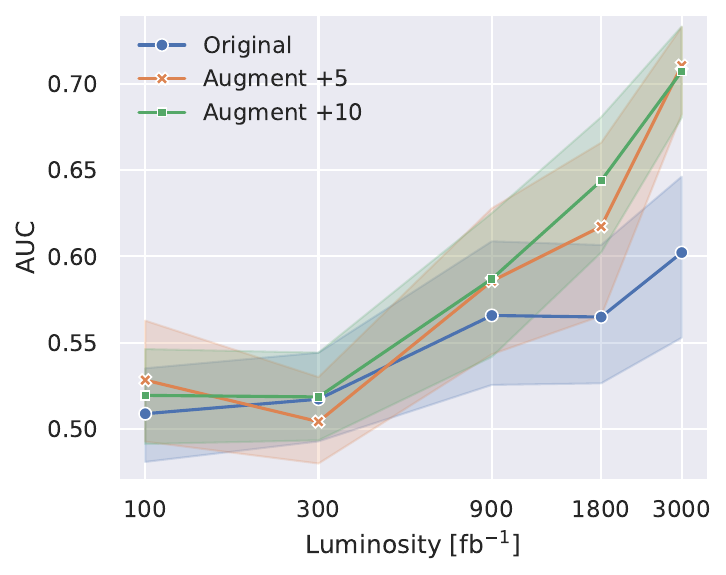}
                \caption{CNN with lepton information}
            \end{subfigure}
            \hfill
            \begin{subfigure}[b]{0.475\textwidth}
                \centering
                \includegraphics[width=\textwidth]{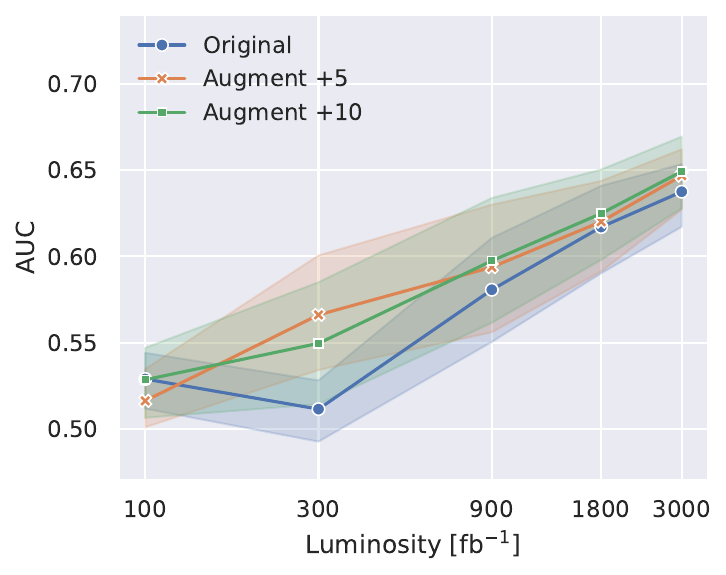}
                \caption{CNN without lepton information}
            \end{subfigure}
            \\[0.75cm]
            \begin{subfigure}[b]{0.475\textwidth}
                \centering
                \includegraphics[width=\textwidth]{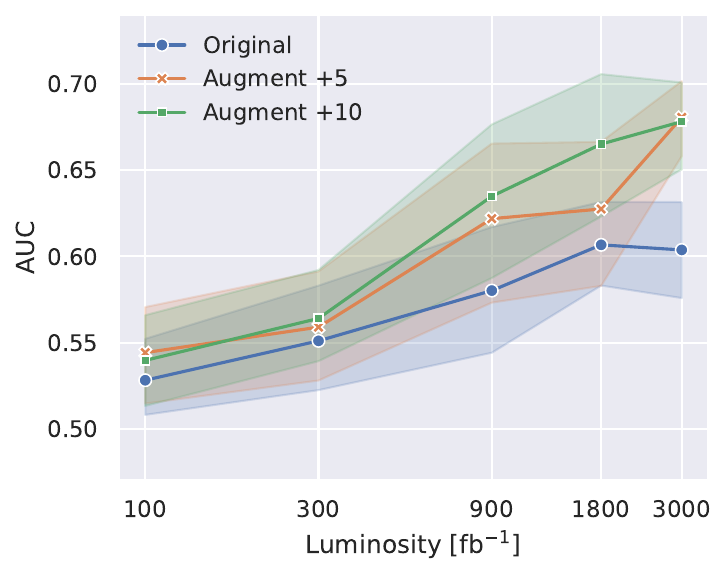}
                \caption{ParT with lepton information}
            \end{subfigure}
            \hfill
            \begin{subfigure}[b]{0.475\textwidth}
                \centering
                \includegraphics[width=\textwidth]{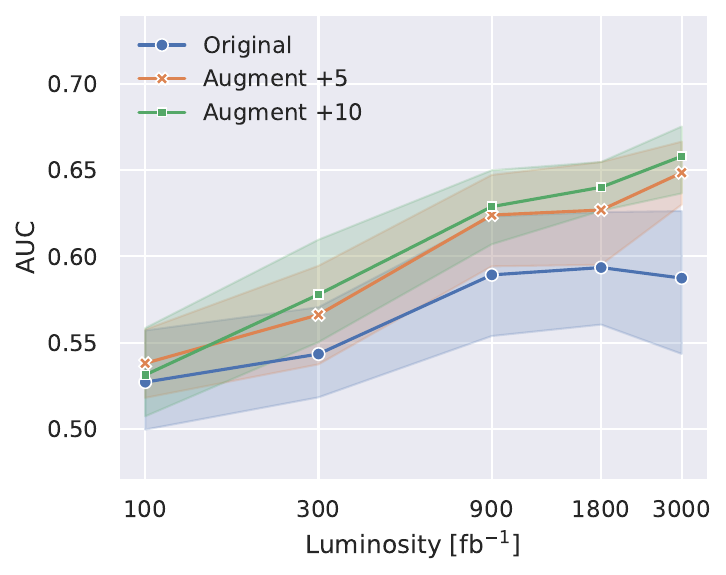}
                \caption{ParT without lepton information}
            \end{subfigure}
        
            \caption{Test AUC of CNN and ParT models trained directly on $H \to ZZ \to 4\ell$ events with and without $\phi$-shifting augmentation. Each point represents the mean AUC over ten training seeds, and the shaded bands indicate one standard deviation.}
			\label{fig:result_zz4l_aug}
        \end{figure}

		To explore possible mitigation strategies, we apply the $\phi$-shifting augmentation technique to expand the effective dataset. The corresponding results are shown in figure~\ref{fig:result_zz4l_aug}. Although data augmentation provides a small improvement in AUC, the overall enhancement remains limited due to the extreme rarity of $H \to ZZ \to 4\ell$ events.

		Overall, the training in this channel is clearly limited by the small number of training events. The resulting training curves exhibit large fluctuations and no consistent improvement for luminosities below $3000~\mathrm{fb}^{-1}$. Even with the application of data augmentation, the performance gain remains marginal. These results indicate that direct weakly supervised training on $H \to ZZ \to 4\ell$ is strongly constrained by statistical limitations and thus less reliable when considered in isolation.


    \subsection{Transferability across Higgs decay channels}
    \label{sub:transferability_across_higgs_decay_channels}

        As demonstrated in section~\ref{sub:performance_on_h_to_zz_to_4l}, direct weakly supervised training on $H \to ZZ \to 4\ell$ is strongly limited by the small event yield. Even with data augmentation, the improvement remains modest. To overcome this limitation, we explore a transfer learning strategy in which models trained on the high-statistics $H \to \gamma\gamma$ dataset are applied to other low-statistics decay channels, including $H \to ZZ \to 4\ell$ and $H \to Z\gamma \to 2\ell\gamma$.

        \begin{figure}[htbp]
            \centering
            \begin{subfigure}[b]{0.475\textwidth}
                \centering
                \includegraphics[width=\textwidth]{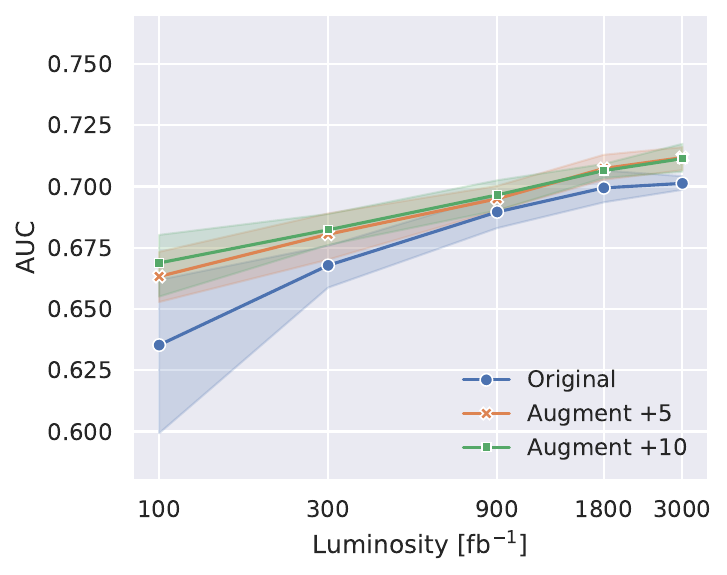}
                \caption{CNN transferred to $H \to ZZ \to 4\ell$}
                \label{fig:transfer_CNN_ZZ}
            \end{subfigure}
            \hfill
            \begin{subfigure}[b]{0.475\textwidth}
                \centering
                \includegraphics[width=\textwidth]{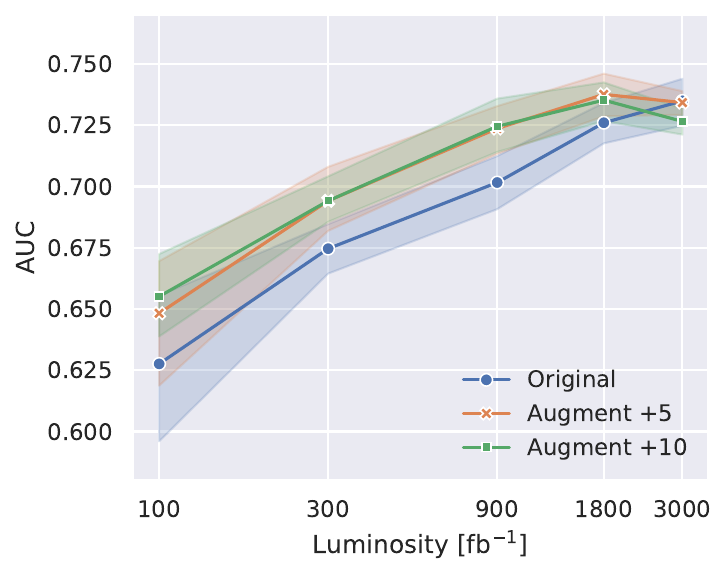}
                \caption{ParT transferred to $H \to ZZ \to 4\ell$}
                \label{fig:transfer_ParT_ZZ}
            \end{subfigure}
            \\[0.5cm]
            \begin{subfigure}[b]{0.475\textwidth}
                \centering
                \includegraphics[width=\textwidth]{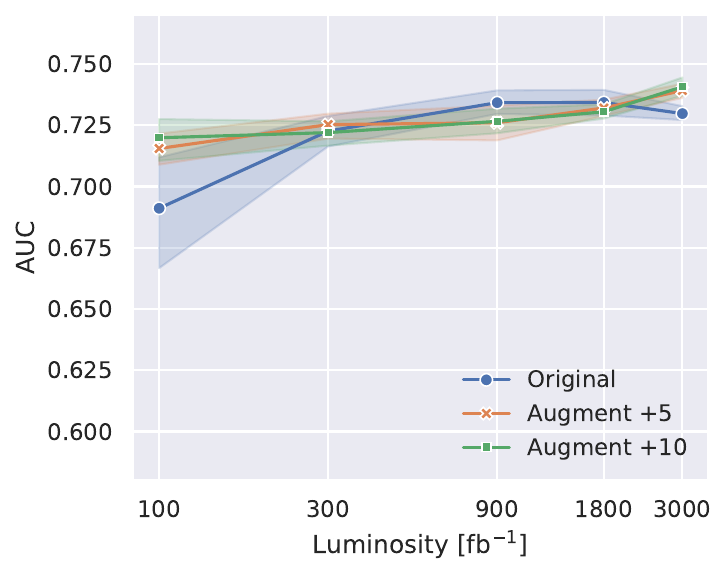}
                \caption{CNN transferred to $H \to Z\gamma \to 2\ell\gamma$}
                \label{fig:transfer_CNN_Zgamma}
            \end{subfigure}
            \hfill
            \begin{subfigure}[b]{0.475\textwidth}
                \centering
                \includegraphics[width=\textwidth]{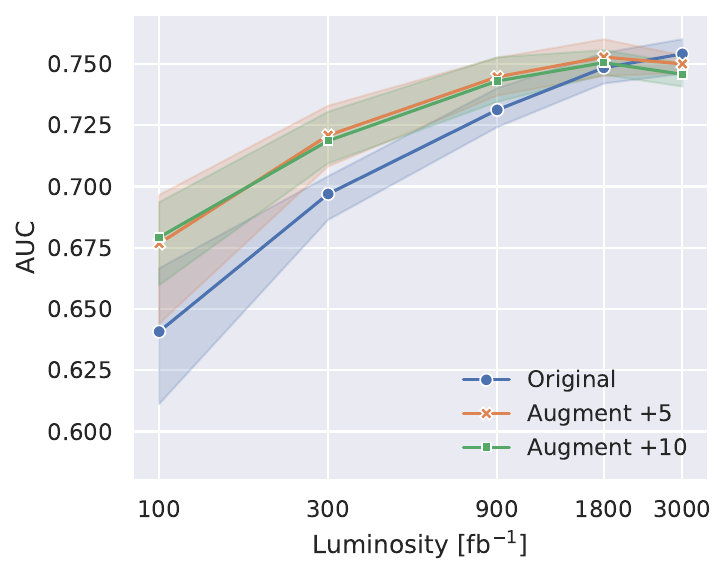}
                \caption{ParT transferred to $H \to Z\gamma \to 2\ell\gamma$}
                \label{fig:transfer_ParT_Zgamma}
            \end{subfigure}

            \caption{Transfer performance of CNN and ParT models pretrained on $H \to \gamma\gamma$ (excluding decay-product information) and evaluated on other Higgs decay channels. Each point represents the mean AUC over ten training seeds, with shaded bands indicating one standard deviation.}
            \label{fig:transfer_results}
        \end{figure}

		Figure~\ref{fig:transfer_results} illustrates the transferability of models trained on $H \to \gamma\gamma$ events to the other channels under consistent conditions, where decay-product information is removed from both the training and evaluation datasets. 

		When applied to $H \to ZZ \to 4\ell$, as shown in figures~\ref{fig:transfer_CNN_ZZ} and \ref{fig:transfer_ParT_ZZ}, both CNN and ParT retain strong discrimination power with the only modest degradation, typically within 0.05 in AUC compared to figures~\ref{fig:result_diphoton_aug-exCNN} and \ref{fig:result_diphoton_aug-exParT}, highlighting a promising level of generalization. This indicates that models trained solely on particle-flow observables, such as calorimeter and tracking data, can retain discriminative power across distinct Higgs decay modes.

		Quantitatively, when compared with the direct training results shown in figure~\ref{fig:CNN_ParT_wo_lepton}, the performance with data augmentation achieved at $\mathcal{L} = 100~\mathrm{fb}^{-1}$ in the transferred models corresponds to the level obtained at approximately $\mathcal{L} = 9,000 - 18,000~\mathrm{fb}^{-1}$ for models trained directly on the $H \to ZZ \to 4\ell$ data. This represents an effective gain of more than two orders of magnitude in statistical efficiency.

        Extending this approach to the $H \to Z\gamma \to 2\ell\gamma$ channel yields even stronger results shown in figures~\ref{fig:transfer_CNN_Zgamma} and \ref{fig:transfer_ParT_Zgamma}. The transferred CNN and ParT models achieve slightly higher AUCs than those obtained on $ZZ$ transfer tests, and their fluctuations across random seeds are smaller. This improvement is attributed to the intermediate event topology of $Z\gamma$, which shares both the hadronic characteristics of $H \to \gamma\gamma$ and the leptonic features of $H \to ZZ \to 4\ell$. As a result, the domain gap between the training and evaluation channels is reduced, enabling better feature reuse and more stable transfer performance.

		These findings demonstrate that in cases where the Higgs decay products do not participate in strong interactions, decay-agnostic classifiers can be trained in one Higgs decay channel ($H \to \gamma\gamma$) and successfully applied to others ($H \to ZZ, Z\gamma$), substantially enhancing the practical utility of the CWoLa framework in data-limited scenarios. Moreover, when assisted with data augmentation, transfer learning further strengthens the robustness and flexibility of weakly supervised classification strategies.



\section{Conclusions}
\label{sec:conclusions}

	In this work, we have applied the CWoLa framework to distinguish between VBF and GGF Higgs boson production mechanisms, using purely experimental data. Deep learning models based on CNNs and ParTs were trained using simulated datasets under various configurations, allowing us to systematically investigate the effects of data representation, augmentation, and transferability across Higgs decay channels.

	We first focused on the $H \to \gamma\gamma$ channel. The results demonstrate that the CWoLa framework effectively captures the differences between VBF and GGF processes even without explicit labels. Moreover, we found that the inclusion of photon information offers only marginal improvement. The hadronic activity alone contains sufficient discriminative information for the production mechanism. This finding indicates that the trained networks primarily rely on global event patterns rather than specific decay product information, revealing that the models could be agnostic to Higgs decay modes.

	To address limited training statistics at low luminosities, we implemented $\phi$-shifting data augmentation. The augmentation improved performance in the low-data regime, enhancing both the mean AUC and the training stability. At higher luminosities, where the dataset is already sufficient, the augmentation effect saturates, confirming its role as an effective regularization and variance-reduction technique in weakly supervised training.

	To further test the decay-channel independence of the learned representations, we extended our study to additional Higgs decay channels.  Due to its extremely small branching ratio, the $H \to ZZ \to 4\ell$ channel suffers from severe data scarcity, and even with data augmentation, the achievable performance remains limited at realistic luminosities. However, when models pretrained on $H \to \gamma\gamma$ samples are evaluated on $H \to ZZ \to 4\ell$ events, with decay-product information removed for both channels, a remarkable improvement is observed. The transferred models retain strong discriminative power, achieving AUC values comparable to those obtained from direct training with nearly two orders of magnitude higher effective luminosity. Furthermore, this transferability can be extended to the $H \to Z\gamma \to 2\ell\gamma$ channel, demonstrating that the learned features generalize well across different Higgs decay modes. These results highlight the potential of CWoLa-trained classifiers to be effectively reused in data-limited scenarios.

	Overall, this study demonstrates that the CWoLa framework, combined with modern deep learning architectures, can successfully distinguish Higgs production mechanisms without relying on explicit labeling. The combination of data augmentation and transfer learning provides practical strategies to enhance model performance in limited-data regimes. These results highlight the potential for applying weakly supervised learning techniques to real LHC analyses, where the data are scarce and transferability across decay modes is essential. Future work could extend this approach to additional Higgs decay channels and explore its integration into experimental workflows.
	

\appendix

\section{Hyperparameters of Particle Transformer} \label{appendix:A}

The architecture of the Particle Transformer used in this work is based on the original design introduced in \cite{Qu:2022mxj}, with several modifications to better suit the characteristics of our dataset. Unlike the dataset used in \cite{Qu:2022mxj}, which contains a broad set of input features derived from fully reconstructed particles, our data consists of three heterogeneous sources: calorimeter towers, reconstructed tracks, and decay-product information. Consequently, some architectural components are adapted accordingly.

In particular, we omit the interaction embedding used in the original implementation, as discussed in section~\ref{sub:neural_network_architectures_and_training_strategy}. The final configuration used in this study is outlined below:
\begin{itemize}
    \item \textbf{Particle Embedding:} Input particle features are embedded into a latent space of dimension $d = 16$ using a three-layer multilayer perceptron with hidden dimensions of 16, 64, and 16. Each layer uses the GELU~\cite{DBLP:journals/corr/HendrycksG16} activation function, and layer normalization is applied between layers to ensure stable training.

    \item \textbf{Particle Attention Block:} A single particle attention block is used, with 4 attention heads. The feedforward network consists of two linear layers with 64 and 16 hidden units, respectively. A dropout rate of 0.1 is applied within this block.

    \item \textbf{Class Attention Block:} One class attention block is included, also using 4 attention heads. Its feedforward layers mirror those of the particle attention block (64 and 16 hidden units), but no dropout is applied in this block (dropout rate = 0.0).
\end{itemize}
All remaining components and architectural details not explicitly listed here follow the original implementation in \cite{Qu:2022mxj}.

\acknowledgments

We thank Zong-En Chen and You-Ying Li for their participation at the early stage of this project.
K.F.C. and Y.A.C. are supported in part by the National Science and Technology Council under Grant No. NSTC-112-2112-M-002-026.
C.W.C. and F.Y.H. are supported in part by the National Science and Technology Council under Grant Nos. NSTC-111-2112-M-002-018-MY3 and NSTC114-2112-M-002-020-MY3. 

\bibliographystyle{JHEP}
\bibliography{biblio.bib}

@Article{trans1,
	title={{A Lorentz-equivariant transformer for all of the LHC}},
	author={Johann Brehmer and Víctor Bresó and Pim de Haan and Tilman Plehn and Huilin Qu and Jonas Spinner and Jesse Thaler},
	journal={SciPost Phys.},
	volume={19},
	pages={108},
	year={2025},
	publisher={SciPost},
	doi={10.21468/SciPostPhys.19.4.108},
	url={https://scipost.org/10.21468/SciPostPhys.19.4.108},
}

@inproceedings{Spinner:2024hjm,
    author = "Spinner, Jonas and Bres{\'o}, Victor and de Haan, Pim and Plehn, Tilman and Thaler, Jesse and Brehmer, Johann",
    title = "{Lorentz-Equivariant Geometric Algebra Transformers for High-Energy Physics}",
    booktitle = "{38th conference on Neural Information Processing Systems}",
    eprint = "2405.14806",
    archivePrefix = "arXiv",
    primaryClass = "physics.data-an",
    reportNumber = "MIT-CTP/5723",
    month = "10",
    year = "2024"
}

@article{Finke:2023veq,
    author = {Finke, Thorben and Kr{\"a}mer, Michael and M{\"u}ck, Alexander and T{\"o}nshoff, Jan},
    title = "{Learning the language of QCD jets with transformers}",
    eprint = "2303.07364",
    archivePrefix = "arXiv",
    primaryClass = "hep-ph",
    doi = "10.1007/JHEP06(2023)184",
    journal = "JHEP",
    volume = "06",
    pages = "184",
    year = "2023"
}

@inproceedings{Li:2023xhj,
    author = "Li, Anni and Krishnamohan, Venkat and Kansal, Raghav and Sen, Rounak and Tsan, Steven and Zhang, Zhaoyu and Duarte, Javier",
    title = "{Induced Generative Adversarial Particle Transformers}",
    booktitle = "{37th Conference on Neural Information Processing Systems}",
    eprint = "2312.04757",
    archivePrefix = "arXiv",
    primaryClass = "hep-ex",
    reportNumber = "FERMILAB-CONF-23-751-CMS-PPD",
    month = "12",
    year = "2023"
}

@inproceedings{Wang:2024rup,
    author = "Wang, Aaron and Gandrakota, Abhijith and Ngadiuba, Jennifer and Sahu, Vivekanand and Bhatnagar, Priyansh and Khoda, Elham E. and Duarte, Javier",
    title = "{Interpreting Transformers for Jet Tagging}",
    eprint = "2412.03673",
    archivePrefix = "arXiv",
    primaryClass = "hep-ph",
    month = "12",
    year = "2024"
}

@article{deOliveira:2015xxd,
    author = "de Oliveira, Luke and Kagan, Michael and Mackey, Lester and Nachman, Benjamin and Schwartzman, Ariel",
    title = "{Jet-images {\textemdash} deep learning edition}",
    eprint = "1511.05190",
    archivePrefix = "arXiv",
    primaryClass = "hep-ph",
    doi = "10.1007/JHEP07(2016)069",
    journal = "JHEP",
    volume = "07",
    pages = "069",
    year = "2016"
}

@article{cnn2,
    author = "Komiske, Patrick T. and Metodiev, Eric M. and Schwartz, Matthew D.",
    title = "{Deep learning in color: towards automated quark/gluon jet discrimination}",
    eprint = "1612.01551",
    archivePrefix = "arXiv",
    primaryClass = "hep-ph",
    reportNumber = "MIT-CTP-4866, MIT-CTP 4866",
    doi = "10.1007/JHEP01(2017)110",
    journal = "JHEP",
    volume = "01",
    pages = "110",
    year = "2017"
}

@article{Kasieczka:2017nvn,
    author = "Kasieczka, Gregor and Plehn, Tilman and Russell, Michael and Schell, Torben",
    title = "{Deep-learning Top Taggers or The End of QCD?}",
    eprint = "1701.08784",
    archivePrefix = "arXiv",
    primaryClass = "hep-ph",
    reportNumber = "MCNET-17-07",
    doi = "10.1007/JHEP05(2017)006",
    journal = "JHEP",
    volume = "05",
    pages = "006",
    year = "2017"
}

@article{Macaluso:2018tck,
    author = "Macaluso, Sebastian and Shih, David",
    title = "{Pulling Out All the Tops with Computer Vision and Deep Learning}",
    eprint = "1803.00107",
    archivePrefix = "arXiv",
    primaryClass = "hep-ph",
    doi = "10.1007/JHEP10(2018)121",
    journal = "JHEP",
    volume = "10",
    pages = "121",
    year = "2018"
}

@Article{cnn5,
author={Lee, Jason Sang Hun
and Park, Inkyu
and Watson, Ian James
and Yang, Seungjin},
title={Quark-Gluon Jet Discrimination Using Convolutional Neural Networks},
journal={Journal of the Korean Physical Society},
year={2019},
month={Feb},
day={01},
volume={74},
number={3},
pages={219-223},
issn={1976-8524},
doi={10.3938/jkps.74.219},
url={https://doi.org/10.3938/jkps.74.219}
}

@article{Alwall:2014hca,
    author = "Alwall, J. and Frederix, R. and Frixione, S. and Hirschi, V. and Maltoni, F. and Mattelaer, O. and Shao, H. -S. and Stelzer, T. and Torrielli, P. and Zaro, M.",
    title = "{The automated computation of tree-level and next-to-leading order differential cross sections, and their matching to parton shower simulations}",
    eprint = "1405.0301",
    archivePrefix = "arXiv",
    primaryClass = "hep-ph",
    reportNumber = "CERN-PH-TH-2014-064, CP3-14-18, LPN14-066, MCNET-14-09, ZU-TH-14-14",
    doi = "10.1007/JHEP07(2014)079",
    journal = "JHEP",
    volume = "07",
    pages = "079",
    year = "2014"
}

@article{Sjostrand:2014zea,
	archiveprefix = {arXiv},
	author = {Sj\"ostrand, Torbj\"orn and Ask, Stefan and Christiansen, Jesper R. and Corke, Richard and Desai, Nishita and Ilten, Philip and Mrenna, Stephen and Prestel, Stefan and Rasmussen, Christine O. and Skands, Peter Z.},
	doi = {10.1016/j.cpc.2015.01.024},
	eprint = {1410.3012},
	journal = {Comput. Phys. Commun.},
	pages = {159--177},
	primaryclass = {hep-ph},
	reportnumber = {LU-TP-14-36, MCNET-14-22, CERN-PH-TH-2014-190, FERMILAB-PUB-14-316-CD, DESY-14-178, SLAC-PUB-16122},
	title = {{An introduction to PYTHIA 8.2}},
	volume = {191},
	year = {2015}
}

@article{deFavereau:2013fsa,
      author         = "de Favereau, J. and Delaere, C. and Demin, P. and
                        Giammanco, A. and Lemaître, V. and Mertens, A. and
                        Selvaggi, M.",
      title          = "{DELPHES 3, A modular framework for fast simulation of a
                        generic collider experiment}",
      collaboration  = "DELPHES 3",
      journal        = "JHEP",
      volume         = "02",
      year           = "2014",
      pages          = "057",
      doi            = "10.1007/JHEP02(2014)057",
      eprint         = "1307.6346",
      archivePrefix  = "arXiv",
      primaryClass   = "hep-ex",
      SLACcitation   = "%%CITATION = ARXIV:1307.6346;%%"
}

@article{Cacciari:2011ma,
	archiveprefix = {arXiv},
	author = {Cacciari, Matteo and Salam, Gavin P. and Soyez, Gregory},
	doi = {10.1140/epjc/s10052-012-1896-2},
	eprint = {1111.6097},
	journal = {Eur. Phys. J. C},
	pages = {1896},
	primaryclass = {hep-ph},
	reportnumber = {CERN-PH-TH-2011-297},
	title = {{FastJet User Manual}},
	volume = {72},
	year = {2012}
}

@article{Cacciari:2008gp,
	archiveprefix = {arXiv},
	author = {Cacciari, Matteo and Salam, Gavin P. and Soyez, Gregory},
	doi = {10.1088/1126-6708/2008/04/063},
	eprint = {0802.1189},
	journal = {JHEP},
	pages = {063},
	primaryclass = {hep-ph},
	reportnumber = {LPTHE-07-03},
	title = {{The anti-$k_t$ jet clustering algorithm}},
	volume = {04},
	year = {2008}
}

@article{LHCHiggsCrossSectionWorkingGroup:2016ypw,
    author = "de Florian, D. and others",
    collaboration = "LHC Higgs Cross Section Working Group",
    title = "{Handbook of LHC Higgs Cross Sections: 4. Deciphering the Nature of the Higgs Sector}",
    eprint = "1610.07922",
    archivePrefix = "arXiv",
    primaryClass = "hep-ph",
    reportNumber = "CERN-2017-002-M, CERN-2017-002",
    doi = "10.23731/CYRM-2017-002",
    journal = "CERN Yellow Rep. Monogr.",
    volume = "2",
    pages = "1--869",
    year = "2017"
}

@article{Chen:2024nvc,
    author = "Chen, Zong-En and Chiang, Cheng-Wei and Hsieh, Feng-Yang",
    title = "{Improving the performance of weak supervision searches using data augmentation}",
    eprint = "2412.00198",
    archivePrefix = "arXiv",
    primaryClass = "hep-ph",
    doi = "10.1007/JHEP09(2025)169",
    journal = "JHEP",
    volume = "09",
    pages = "169",
    year = "2025"
}

@article{ATLAS:2012yve,
    author = "Aad, Georges and others",
    collaboration = "ATLAS",
    title = "{Observation of a new particle in the search for the Standard Model Higgs boson with the ATLAS detector at the LHC}",
    eprint = "1207.7214",
    archivePrefix = "arXiv",
    primaryClass = "hep-ex",
    reportNumber = "CERN-PH-EP-2012-218",
    doi = "10.1016/j.physletb.2012.08.020",
    journal = "Phys. Lett. B",
    volume = "716",
    pages = "1--29",
    year = "2012"
}

@article{CMS:2012qbp,
    author = "Chatrchyan, Serguei and others",
    collaboration = "CMS",
    title = "{Observation of a New Boson at a Mass of 125 GeV with the CMS Experiment at the LHC}",
    eprint = "1207.7235",
    archivePrefix = "arXiv",
    primaryClass = "hep-ex",
    reportNumber = "CMS-HIG-12-028, CERN-PH-EP-2012-220",
    doi = "10.1016/j.physletb.2012.08.021",
    journal = "Phys. Lett. B",
    volume = "716",
    pages = "30--61",
    year = "2012"
}

@article{ATLAS:2022vkf,
    author = "Aad, Georges and others",
    collaboration = "ATLAS",
    title = "{A detailed map of Higgs boson interactions by the ATLAS experiment ten years after the discovery}",
    eprint = "2207.00092",
    archivePrefix = "arXiv",
    primaryClass = "hep-ex",
    reportNumber = "CERN-EP-2022-057",
    doi = "10.1038/s41586-022-04893-w",
    journal = "Nature",
    volume = "607",
    number = "7917",
    pages = "52--59",
    year = "2022",
    note = "[Erratum: Nature 612, E24 (2022)]"
}

@article{CMS:2022dwd,
    author = "Tumasyan, Armen and others",
    collaboration = "CMS",
    title = "{A portrait of the Higgs boson by the CMS experiment ten years after the discovery.}",
    eprint = "2207.00043",
    archivePrefix = "arXiv",
    primaryClass = "hep-ex",
    reportNumber = "CMS-HIG-22-001, CERN-EP-2022-039",
    doi = "10.1038/s41586-022-04892-x",
    journal = "Nature",
    volume = "607",
    number = "7917",
    pages = "60--68",
    year = "2022",
    note = "[Erratum: Nature 623, (2023)]"
}

@article{Chiang:2022lsn,
    author = "Chiang, Cheng-Wei and Shih, David and Wei, Shang-Fu",
    title = "{VBF vs. GGF Higgs with Full-Event Deep Learning: Towards a Decay-Agnostic Tagger}",
    eprint = "2209.05518",
    archivePrefix = "arXiv",
    primaryClass = "hep-ph",
    doi = "10.1103/PhysRevD.107.016014",
    journal = "Phys. Rev. D",
    volume = "107",
    number = "1",
    pages = "016014",
    year = "2023"
}

@article{Metodiev:2017vrx,
    author = "Metodiev, Eric M. and Nachman, Benjamin and Thaler, Jesse",
    title = "{Classification without labels: Learning from mixed samples in high energy physics}",
    eprint = "1708.02949",
    archivePrefix = "arXiv",
    primaryClass = "hep-ph",
    reportNumber = "MIT--CTP-4922",
    doi = "10.1007/JHEP10(2017)174",
    journal = "JHEP",
    volume = "10",
    pages = "174",
    year = "2017"
}

@article{Chung:2020ysf,
    author = "Chung, Yi-Lun and Hsu, Shih-Chieh and Nachman, Benjamin",
    title = "{Disentangling Boosted Higgs Boson Production Modes with Machine Learning}",
    eprint = "2009.05930",
    archivePrefix = "arXiv",
    primaryClass = "hep-ph",
    doi = "10.1088/1748-0221/16/07/P07002",
    journal = "JINST",
    volume = "16",
    pages = "P07002",
    year = "2021"
}

@article{Builtjes:2022usj,
    author = "Builtjes, Luc and Caron, Sascha and Moskvitina, Polina and Nellist, Clara and de Austri, Roberto Ruiz and Verheyen, Rob and Zhang, Zhongyi",
    title = "{Attention to the strengths of physical interactions: Transformer and graph-based event classification for particle physics experiments}",
    eprint = "2211.05143",
    archivePrefix = "arXiv",
    primaryClass = "hep-ph",
    doi = "10.21468/SciPostPhys.19.1.028",
    journal = "SciPost Phys.",
    volume = "19",
    number = "1",
    pages = "028",
    year = "2025"
}

@article{Auricchio:2023syb,
    author = "Auricchio, Silvia and Cirotto, Francesco and Giannini, Antonio",
    title = "{VBF Event Classification with Recurrent Neural Networks at ATLAS{\textquoteright}s LHC Experiment}",
    doi = "10.3390/app13053282",
    journal = "Appl. Sciences",
    volume = "13",
    number = "5",
    pages = "3282",
    year = "2023"
}

@article{Qu:2022mxj,
    author = "Qu, Huilin and Li, Congqiao and Qian, Sitian",
    title = "{Particle Transformer for Jet Tagging}",
    eprint = "2202.03772",
    archivePrefix = "arXiv",
    primaryClass = "hep-ph",
    month = "2",
    year = "2022"
}

@article{Li:2019ufu,
    author = "Li, Gexing and Li, Zhao and Wang, Yan and Wang, Yefan",
    title = "{Improving the measurement of the Higgs boson-gluon coupling using convolutional neural networks at $e^+e^-$ colliders}",
    eprint = "1901.09391",
    archivePrefix = "arXiv",
    primaryClass = "hep-ph",
    doi = "10.1103/PhysRevD.100.116013",
    journal = "Phys. Rev. D",
    volume = "100",
    number = "11",
    pages = "116013",
    year = "2019"
}

@article{ATLAS:2018jvf,
    author = "Aaboud, Morad and others",
    collaboration = "ATLAS",
    title = "{Search for Higgs bosons produced via vector-boson fusion and decaying into bottom quark pairs in $\sqrt{s} = 13$ $\mathrm{TeV}$ $pp$ collisions with the ATLAS detector}",
    eprint = "1807.08639",
    archivePrefix = "arXiv",
    primaryClass = "hep-ex",
    reportNumber = "CERN-EP-2018-140",
    doi = "10.1103/PhysRevD.98.052003",
    journal = "Phys. Rev. D",
    volume = "98",
    number = "5",
    pages = "052003",
    year = "2018"
}

@article{CMS:2023tfj,
    author = "Hayrapetyan, Aram and others",
    collaboration = "CMS",
    title = "{Measurement of the Higgs boson production via vector boson fusion and its decay into bottom quarks in proton-proton collisions at $ \sqrt{s} $ = 13 TeV}",
    eprint = "2308.01253",
    archivePrefix = "arXiv",
    primaryClass = "hep-ex",
    reportNumber = "CMS-HIG-22-009, CERN-EP-2023-110",
    doi = "10.1007/JHEP01(2024)173",
    journal = "JHEP",
    volume = "01",
    pages = "173",
    year = "2024"
}

@article{Beauchesne:2023vie,
    author = "Beauchesne, Hugues and Chen, Zong-En and Chiang, Cheng-Wei",
    title = "{Improving the performance of weak supervision searches using transfer and meta-learning}",
    eprint = "2312.06152",
    archivePrefix = "arXiv",
    primaryClass = "hep-ph",
    doi = "10.1007/JHEP02(2024)138",
    journal = "JHEP",
    volume = "02",
    pages = "138",
    year = "2024"
}

@article{Collins:2018epr,
    author = "Collins, Jack H. and Howe, Kiel and Nachman, Benjamin",
    title = "{Anomaly Detection for Resonant New Physics with Machine Learning}",
    eprint = "1805.02664",
    archivePrefix = "arXiv",
    primaryClass = "hep-ph",
    reportNumber = "FERMILAB-PUB-18-180-T",
    doi = "10.1103/PhysRevLett.121.241803",
    journal = "Phys. Rev. Lett.",
    volume = "121",
    number = "24",
    pages = "241803",
    year = "2018"
}

@article{Collins:2019jip,
    author = "Collins, Jack H. and Howe, Kiel and Nachman, Benjamin",
    title = "{Extending the search for new resonances with machine learning}",
    eprint = "1902.02634",
    archivePrefix = "arXiv",
    primaryClass = "hep-ph",
    reportNumber = "FERMILAB-PUB-18-733-T",
    doi = "10.1103/PhysRevD.99.014038",
    journal = "Phys. Rev. D",
    volume = "99",
    number = "1",
    pages = "014038",
    year = "2019"
}

@article{DBLP:journals/corr/abs-1912-01703,
  author       = {Adam Paszke and
                  Sam Gross and
                  Francisco Massa and
                  Adam Lerer and
                  James Bradbury and
                  Gregory Chanan and
                  Trevor Killeen and
                  Zeming Lin and
                  Natalia Gimelshein and
                  Luca Antiga and
                  Alban Desmaison and
                  Andreas K{\"{o}}pf and
                  Edward Z. Yang and
                  Zach DeVito and
                  Martin Raison and
                  Alykhan Tejani and
                  Sasank Chilamkurthy and
                  Benoit Steiner and
                  Lu Fang and
                  Junjie Bai and
                  Soumith Chintala},
  title        = {PyTorch: An Imperative Style, High-Performance Deep Learning Library},
  journal      = {CoRR},
  volume       = {abs/1912.01703},
  year         = {2019},
  url          = {http://arxiv.org/abs/1912.01703},
  eprinttype    = {arXiv},
  eprint       = {1912.01703},
  bibsource    = {dblp computer science bibliography, https://dblp.org}
}

@article{Gallicchio:2011xq,
    author = "Gallicchio, Jason and Schwartz, Matthew D.",
    title = "{Quark and Gluon Tagging at the LHC}",
    eprint = "1106.3076",
    archivePrefix = "arXiv",
    primaryClass = "hep-ph",
    doi = "10.1103/PhysRevLett.107.172001",
    journal = "Phys. Rev. Lett.",
    volume = "107",
    pages = "172001",
    year = "2011"
}

@inproceedings{DBLP:journals/corr/KingmaB14,
  author       = {Diederik P. Kingma and
                  Jimmy Ba},
  editor       = {Yoshua Bengio and
                  Yann LeCun},
  title        = {Adam: {A} Method for Stochastic Optimization},
  booktitle    = {3rd International Conference on Learning Representations, {ICLR} 2015,
                  San Diego, CA, USA, May 7-9, 2015, Conference Track Proceedings},
  year         = {2015},
  url          = {http://arxiv.org/abs/1412.6980},
  bibsource    = {dblp computer science bibliography, https://dblp.org}
}

@article{Lee:2019ssx,
    author = "Lee, Jason Sang Hun and Lee, Sang Man and Lee, Yunjae and Park, Inkyu and Watson, Ian James and Yang, Seungjin",
    title = "{Quark Gluon Jet Discrimination with Weakly Supervised Learning}",
    eprint = "2012.02540",
    archivePrefix = "arXiv",
    primaryClass = "hep-ph",
    doi = "10.3938/jkps.75.652",
    journal = "J. Korean Phys. Soc.",
    volume = "75",
    number = "9",
    pages = "652--659",
    year = "2019"
}

@article{Dolan:2023abg,
    author = "Dolan, Matthew J. and Gargalionis, John and Ore, Ayodele",
    title = "{Quark-versus-gluon tagging in CMS Open Data with CWoLa and TopicFlow}",
    eprint = "2312.03434",
    archivePrefix = "arXiv",
    primaryClass = "hep-ph",
    doi = "10.1007/JHEP08(2025)024",
    journal = "JHEP",
    volume = "08",
    pages = "024",
    year = "2025"
}

@article{Greljo:2015sla,
    author = "Greljo, Admir and Isidori, Gino and Lindert, Jonas M. and Marzocca, David",
    title = "{Pseudo-observables in electroweak Higgs production}",
    eprint = "1512.06135",
    archivePrefix = "arXiv",
    primaryClass = "hep-ph",
    reportNumber = "ZU-TH-47-15",
    doi = "10.1140/epjc/s10052-016-4000-5",
    journal = "Eur. Phys. J. C",
    volume = "76",
    number = "3",
    pages = "158",
    year = "2016"
}

@article{DBLP:journals/corr/abs-1912-08001,
  author       = {Marouen Baalouch and
                  Maxime Defurne and
                  Jean{-}Philippe Poli and
                  No{\"{e}}lie Cherrier},
  title        = {Sim-to-Real Domain Adaptation For High Energy Physics},
  journal      = {CoRR},
  volume       = {abs/1912.08001},
  year         = {2019},
  url          = {http://arxiv.org/abs/1912.08001},
  eprinttype    = {arXiv},
  eprint       = {1912.08001},
  bibsource    = {dblp computer science bibliography, https://dblp.org}
}

@article{Gavranovic:2023oam,
    author = "Gavranovi{\v{c}}, Jan and Ker{\v{s}}evan, Borut Paul",
    title = "{Systematic evaluation of generative machine learning capability to simulate distributions of observables at the large hadron collider}",
    eprint = "2310.08994",
    archivePrefix = "arXiv",
    primaryClass = "hep-ph",
    doi = "10.1140/epjc/s10052-024-13284-6",
    journal = "Eur. Phys. J. C",
    volume = "84",
    number = "9",
    pages = "911",
    year = "2024"
}

@article{ATLAS:2020wny,
    author = "Aad, Georges and others",
    collaboration = "ATLAS",
    title = "{Measurements of the Higgs boson inclusive and differential fiducial cross sections in the 4$\ell $ decay channel at $\sqrt{s}$ = 13 TeV}",
    eprint = "2004.03969",
    archivePrefix = "arXiv",
    primaryClass = "hep-ex",
    reportNumber = "CERN-EP-2020-035",
    doi = "10.1140/epjc/s10052-020-8223-0",
    journal = "Eur. Phys. J. C",
    volume = "80",
    number = "10",
    pages = "942",
    year = "2020"
}

@article{Feickert:2021ajf,
    author = "Feickert, Matthew and Nachman, Benjamin",
    title = "{A Living Review of Machine Learning for Particle Physics}",
    eprint = "2102.02770",
    archivePrefix = "arXiv",
    primaryClass = "hep-ph",
    month = "2",
    year = "2021"
}

@article{Ball:2012cx,
    author = "Ball, Richard D. and others",
    title = "{Parton distributions with LHC data}",
    eprint = "1207.1303",
    archivePrefix = "arXiv",
    primaryClass = "hep-ph",
    reportNumber = "EDINBURGH-2012-08, IFUM-FT-997, FR-PHENO-2012-014, RWTH-TTK-12-25, CERN-PH-TH-2012-037, SFB-CPP-12-47",
    doi = "10.1016/j.nuclphysb.2012.10.003",
    journal = "Nucl. Phys. B",
    volume = "867",
    pages = "244--289",
    year = "2013"
}

@misc{hsieh_2025_17628988,
  author       = {Hsieh, Feng-Yang},
  title        = {Dataset for "Higgs Production Classifier using
                   Weak Supervision"
                  },
  month        = nov,
  year         = 2025,
  publisher    = {Zenodo},
  doi          = {10.5281/zenodo.17628988},
  url          = {https://doi.org/10.5281/zenodo.17628988},
}

@article{DBLP:journals/corr/abs-1803-08375,
  author       = {Abien Fred Agarap},
  title        = {Deep Learning using Rectified Linear Units (ReLU)},
  journal      = {CoRR},
  volume       = {abs/1803.08375},
  year         = {2018},
  url          = {http://arxiv.org/abs/1803.08375},
  eprinttype    = {arXiv},
  eprint       = {1803.08375},
  bibsource    = {dblp computer science bibliography, https://dblp.org}
}

@article{DBLP:journals/corr/IoffeS15,
  author       = {Sergey Ioffe and
                  Christian Szegedy},
  title        = {Batch Normalization: Accelerating Deep Network Training by Reducing
                  Internal Covariate Shift},
  journal      = {CoRR},
  volume       = {abs/1502.03167},
  year         = {2015},
  url          = {http://arxiv.org/abs/1502.03167},
  eprinttype    = {arXiv},
  eprint       = {1502.03167},
  bibsource    = {dblp computer science bibliography, https://dblp.org}
}

@article{DBLP:journals/corr/HeZRS15,
  author       = {Kaiming He and
                  Xiangyu Zhang and
                  Shaoqing Ren and
                  Jian Sun},
  title        = {Deep Residual Learning for Image Recognition},
  journal      = {CoRR},
  volume       = {abs/1512.03385},
  year         = {2015},
  url          = {http://arxiv.org/abs/1512.03385},
  eprinttype    = {arXiv},
  eprint       = {1512.03385},
  bibsource    = {dblp computer science bibliography, https://dblp.org}
}

@article{DBLP:journals/corr/HendrycksG16,
  author       = {Dan Hendrycks and
                  Kevin Gimpel},
  title        = {Bridging Nonlinearities and Stochastic Regularizers with Gaussian
                  Error Linear Units},
  journal      = {CoRR},
  volume       = {abs/1606.08415},
  year         = {2016},
  url          = {http://arxiv.org/abs/1606.08415},
  eprinttype    = {arXiv},
  eprint       = {1606.08415},
  bibsource    = {dblp computer science bibliography, https://dblp.org}
}

\end{document}